\documentclass[twocolumn]{aastex631}

\usepackage{amsmath}
\usepackage{here}

\newcommand{\hahb}{\mathrm{H}\alpha/\mathrm{H}\beta}
\newcommand{\hghb}{\mathrm{H}\gamma/\mathrm{H}\beta}
\newcommand{\hdhb}{\mathrm{H}\delta/\mathrm{H}\beta}
\newcommand{\ha}{\mathrm{H}\alpha}
\newcommand{\hb}{\mathrm{H}\beta}
\newcommand{\hg}{\mathrm{H}\gamma}
\newcommand{\hd}{\mathrm{H}\delta}

\begin{document}

\shorttitle{
Balmer Decrement Anomalies in High-$z$ Galaxies
}

\shortauthors{Yanagisawa et al.}

\title{
Balmer Decrement Anomalies in Galaxies at $z\sim 6$ Found by JWST Observations:\\
Density-Bounded Nebulae or Excited H{\sc i} Clouds?
}

\author[0009-0006-6763-4245]{Hiroto Yanagisawa}
\affiliation{Institute for Cosmic Ray Research, The University of Tokyo, 5-1-5 Kashiwanoha, Kashiwa, Chiba 277-8582, Japan}
\affiliation{Department of Physics, Graduate School of Science, The University of Tokyo, 7-3-1 Hongo, Bunkyo, Tokyo 113-0033, Japan}

\author[0000-0002-1049-6658]{Masami Ouchi}
\affiliation{National Astronomical Observatory of Japan, National Institutes of Natural Sciences, 2-21-1 Osawa, Mitaka, Tokyo 181-8588, Japan}
\affiliation{Institute for Cosmic Ray Research, The University of Tokyo, 5-1-5 Kashiwanoha, Kashiwa, Chiba 277-8582, Japan}
\affiliation{Department of Astronomical Science, SOKENDAI (The Graduate University for Advanced Studies), Osawa 2-21-1, Mitaka, Tokyo, 181-8588, Japan}
\affiliation{Kavli Institute for the Physics and Mathematics of the Universe (WPI), University of Tokyo, Kashiwa, Chiba 277-8583, Japan}

\author[0000-0003-2965-5070]{Kimihiko Nakajima}
\affiliation{National Astronomical Observatory of Japan, National Institutes of Natural Sciences, 2-21-1 Osawa, Mitaka, Tokyo 181-8588, Japan}

\author[0000-0002-1319-3433]{Hidenobu Yajima}
\affiliation{Center for Computational Sciences, University of Tsukuba, Ten-nodai, 1-1-1 Tsukuba, Ibaraki 305-8577, Japan}

\author[0009-0008-0167-5129]{Hiroya Umeda}
\affiliation{Institute for Cosmic Ray Research, The University of Tokyo, 5-1-5 Kashiwanoha, Kashiwa, Chiba 277-8582, Japan}
\affiliation{Department of Physics, Graduate School of Science, The University of Tokyo, 7-3-1 Hongo, Bunkyo, Tokyo 113-0033, Japan}

\author[0000-0002-9850-6290]{Shunsuke Baba}
\affiliation{Institute of Space and Astronautical Science (ISAS), Japan Aerospace Exploration Agency (JAXA),
3-1-1 Yoshinodai, Chuo-ku, Sagamihara, Kanagawa 252-5210, Japan}

\author[0000-0002-6660-9375]{Takao Nakagawa}
\affiliation{Institute of Space and Astronautical Science (ISAS), 
Japan Aerospace Exploration Agency (JAXA), 
3-1-1 Yoshinodai, Chuo-ku, Sagamihara, Kanagawa 252-5210, Japan}

\author[0009-0000-1999-5472]{Minami Nakane}
\affiliation{Institute for Cosmic Ray Research, The University of Tokyo, 5-1-5 Kashiwanoha, Kashiwa, Chiba 277-8582, Japan}
\affiliation{Department of Physics, Graduate School of Science, The University of Tokyo, 7-3-1 Hongo, Bunkyo, Tokyo 113-0033, Japan}

\author{Akinori Matsumoto}
\affiliation{Institute for Cosmic Ray Research, The University of Tokyo, 5-1-5 Kashiwanoha, Kashiwa, Chiba 277-8582, Japan}
\affiliation{Department of Physics, Graduate School of Science, The University of Tokyo, 7-3-1 Hongo, Bunkyo, Tokyo 113-0033, Japan}

\author[0000-0001-9011-7605]{Yoshiaki Ono}
\affiliation{Institute for Cosmic Ray Research, The University of Tokyo, 5-1-5 Kashiwanoha, Kashiwa, Chiba 277-8582, Japan}

\author[0000-0002-6047-430X]{Yuichi Harikane}
\affiliation{Institute for Cosmic Ray Research, The University of Tokyo, 5-1-5 Kashiwanoha, Kashiwa, Chiba 277-8582, Japan}

\author[0000-0001-7730-8634]{Yuki Isobe}
\affiliation{Institute for Cosmic Ray Research, The University of Tokyo, 5-1-5 Kashiwanoha, Kashiwa, Chiba 277-8582, Japan}
\affiliation{Department of Physics, Graduate School of Science, The University of Tokyo, 7-3-1 Hongo, Bunkyo, Tokyo 113-0033, Japan}

\author[0000-0002-5768-8235]{Yi Xu}
\affiliation{Institute for Cosmic Ray Research, The University of Tokyo, 5-1-5 Kashiwanoha, Kashiwa, Chiba 277-8582, Japan}
\affiliation{Department of Astronomy, Graduate School of Science, The University of Tokyo, 7-3-1 Hongo, Bunkyo, Tokyo 113-0033, Japan}

\author[0000-0003-3817-8739]{Yechi Zhang}
\affiliation{National Astronomical Observatory of Japan, National Institutes of Natural Sciences, 2-21-1 Osawa, Mitaka, Tokyo 181-8588, Japan}

\begin{abstract}
We investigate the physical origins of the Balmer decrement anomalies in GS-NDG-9422 (Cameron et al. 2023) and RXCJ2248-ID (Topping et al. 2024) galaxies at $z\sim 6$ whose $\hahb$ values are significantly smaller than $2.7$, the latter of which also shows anomalous $\hghb$ and $\hdhb$ values beyond the errors. Because the anomalous Balmer decrements are not reproduced under the Case B recombination, we explore the nebulae with the optical depths smaller and larger than the Case B recombination by physical modeling. We find two cases quantitatively explaining the anomalies; 1) density-bounded nebulae that are opaque only up to around Ly$\gamma$-Ly8 transitions and 2) ionization-bounded nebulae partly/fully surrounded by optically-thick excited H{\sc i} clouds. The case of 1) produces more H$\beta$ photons via Ly$\gamma$ absorption in the nebulae, requiring fine tuning in optical depth values, while this case helps ionizing photon escape for cosmic reionization. The case of 2) needs the optically-thick excited H{\sc i} clouds with $N_2\simeq 10^{12}-10^{13}$ $\mathrm{cm^{-2}}$, where $N_2$ is the column density of the hydrogen atom with the principal quantum number of $n=2$. Interestingly, the high $N_2$ values qualitatively agree with the recent claims for GS-NDG-9422 with the strong nebular continuum requiring a number of $2s$-state electrons and for RXCJ2248-ID with the dense ionized regions likely coexisting with the optically-thick clouds. While the physical origin of the optically-thick excited H{\sc i} clouds is unclear, these results may suggest gas clouds with excessive collisional excitation 
caused by an amount of accretion and supernovae in the high-$z$ galaxies. 
\end{abstract}


\section{Introduction} \label{sec:intro}
Balmer decrements (i.e., deviations of hydrogen Balmer line ratios from theoretical values)
are known as tracers of dust extinction in galaxies. However, recent observations with James Webb Space Telescope (JWST) have identified anomalous Balmer decrements in high-$z$ galaxies. \cite{Cameron+2023} have reported that GS-NDG-9422 at $z=5.943$ shows $\hahb$ significantly smaller than the theoretical limit with no dust extinction at the $>1\sigma$ level. Similarly, \cite{Topping+2024} have found that RXCJ2248-ID at $z=6.11$ has the anomalously small $\hahb$ value together with the $\hghb$ and $\hdhb$ values significantly larger than the theoretical limits with no dust extinction. These Balmer decrements cannot be explained by the dust extinction, since the dust extinction raises $\hahb$ and reduces $\hghb$ and $\hdhb$. 
Although these studies do not discuss the physical origins of these anomalies quantitatively, there may be an important physical mechanism that changes Balmer line ratios beyond the well-known theoretical limits that are calculated under the Case B recombination corresponding to the optically-thick Lyman series. To understand the Balmer decrement anomalies, our study aims to quantitatively explore the physical origins of the Balmer decrement anomalies, allowing the optical depths smaller and larger than the Case B recombination.

The structure of this paper is as follows. In Section \ref{sec:data}, we present data used in this paper. In Section \ref{sec:explanation}, we explore the possible physical origins of the Balmer decrement anomalies. In Section \ref{sec:discussion}, we present some implications of our results. 
Section \ref{sec:summary} summarizes our results.

\section{Data} \label{sec:data}
In this paper, we investigate two galaxies, GS-NDG-9422 and RXCJ2248-ID, at $z\sim 6$ with the Balmer decrement anomalies reported in the previous studies. We summarize the Balmer line ratios of the GS-NDG-9422 and RXCJ2248-ID in Table \ref{tab:balmer_decrements}.

\begin{deluxetable}{ccc}
    \centering
    \tablecaption{Balmer line ratios of our sample}
    \tablehead{
    \colhead{} & \colhead{GS-NDG-9422} & \colhead{RXCJ2248-ID}
    }
    \startdata
    $\hahb$&$2.65\pm0.03$&$2.547\pm0.022^\dagger$ 
    \\
    $\hghb$&$0.48\pm0.02$&$0.512\pm0.008$\\
    $\hdhb$&$0.25\pm0.03$&$0.283\pm0.012$
    \enddata
    \tablecomments{($\dagger$) The reported flux refers to the narrow component (M. Topping, private communication). }
    \label{tab:balmer_decrements}
\end{deluxetable}

\subsection{GS-NDG-9422} \label{sec:data_gs}
GS-NDG-9422 was observed with JWST NIRSpec via the JWST Advanced Deep Extragalactic Survey (JADES; PID: 1210, PI: Luetzgendorf; \citealt{Bunker+2023a, Eisenstein+2023}). \cite{Cameron+2023} have analysed the spectra and measured fluxes of the rest-UV and optical emission lines. \cite{Cameron+2023} have reported the Balmer line ratios of $\hahb=2.65\pm0.03$, $\hghb=0.48\pm0.02$, and $\hdhb=0.25\pm0.03$. Because all of these flux ratios are calculated from the Prism data covering all of these Balmer emission lines through the one slit position, slit-loss effect is not a cause of the Balmer decrement anomalies. None of these Balmer emission lines show multiple velocity components including a broad component.

\subsection{RXCJ2248-ID} \label{sec:data_rx}
RXCJ2248-ID was observed with JWST NIRSpec in General Observers (GO) program (PID: 2478, PI: Stark). \cite{Topping+2024} have investigated the spectra and presented the Balmer line ratios of $\hahb=2.547\pm0.022$, $\hghb=0.512\pm0.008$, and $\hdhb=0.283\pm0.012$. Because the $\ha$ and $\hb$ emission lines are covered by the single G395M spectrum, the difference of the slit-loss effect is not problematic in the $\hahb$ ratio. Although the $\hg$ and $\hd$ are covered by both G235M and G395M gratings, whether the reported fluxes are taken from G235M or G395M is not clarified in the literature. The $\ha$ line shows a broad component with the broad-to-narrow line flux ratio of 0.22, while the other hydrogen lines do not. The $\hahb$ value listed in Table \ref{tab:balmer_decrements} refers to the narrow-line $\ha$ fluxes (M. Topping, private communication).

\begin{figure}[t]
    \centering
    \includegraphics[width=1\linewidth]{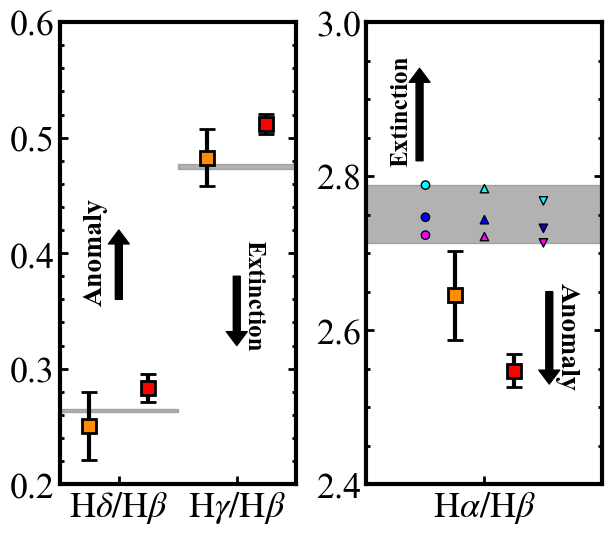}
    \caption{Comparisons of the observed and Case B Balmer line ratios. The orange and red squares represent the values of GS-NDG-9422 \citep{Cameron+2023} and RXCJ2248-ID \citep{Topping+2024}. The gray shaded regions denote the Case B values, where we change the electron temperature from $15,000 \, \mathrm{K}$ to $25,000 \, \mathrm{K}$ and the electron density from $10^{1} \, \mathrm{cm^{-3}}$ to $10^5 \, \mathrm{cm^{-3}}$. In the right panel, the Case B values of a $3\times3$ grid of $T_\mathrm{e}$ and $n_\mathrm{e}$ are shown. The cyan, blue, and magenta points represent the  values of $T_\mathrm{e} = 15,000 \, \mathrm{K}$, $20,000 \, \mathrm{K}$, and $25,000 \, \mathrm{K}$, respectively. The circles, triangles, and inverted triangles show the values of $n_\mathrm{e} = 10^1 \, \mathrm{cm^{-3}}$, $10^3 \, \mathrm{cm^{-3}}$, and $10^5 \, \mathrm{cm^{-3}}$, respectively. The black arrows represent the directions of the effect of the dust extinction and the Balmer decrement anomalies.}
    \label{fig:Balmer_decrements}
\end{figure}

\begin{deluxetable*}{ccc}
    \centering
    \tablecaption{Parameters used in \textsc{Cloudy}}
    \tablehead{
    \colhead{Parameter} & \colhead{Value} & \colhead{Description}
    }
    \startdata
    Blackbody & 20,000, 40,000 & Incident radiation [K] \\
    $\log U$ & $-2, -1$ & Ionization parameter \\
    $\log n$ & 2, 5 & Density of the hydrogen gas [$\mathrm{cm^{-3}}$] \\
    Stop efrac$^{(\dagger)}$ & $-1$ & Stopping criterion \\
    Stop thickness$^{(\dagger\dagger)}$ & $10^{10-20}$ & Stopping criterion [$\mathrm{cm}$] \\
    Geometry & Plane-parallel & Geometry of the gas \\
    Abundance & $0.1$ & Chemical composition of the gas [$Z_\odot$]
    \enddata
    \tablecomments{($\dagger$) When log of the electron fraction is below this value, the calculation stops. ($\dagger\dagger$) When the thickness of the nebula reaches this value, the calculation stops.}
    \label{tab:cloudy_params}
\end{deluxetable*}

\section{Explanation for the Balmer Decrement Anomalies}  \label{sec:explanation}
Here we explore the origin of the Balmer decrement anomalies presented in Section \ref{sec:data}. In Section \ref{sec:caseb}, we compare the anomalies with the values in the Case B recombination, where we define the Case B as optically-thick in the Lyman series and optically-thin in the other hydrogen lines.
We then investigate the two scenarios with optical depths smaller and larger than the one of the Case B recombination that are optically-thin in the Lyman series (Section \ref{sec:hii}) and optically-thick in the Balmer series (Section \ref{sec:optical_depth}), respectively.

\subsection{Comparison with the Case B Recombination} \label{sec:caseb}
Figure \ref{fig:Balmer_decrements} compares the observed and Case B values of Balmer line ratios. The Case B values are calculated with PyNeb (version 1.1.18, \citealt{Luridiana+2015}) for electron temperatures $T_\mathrm{e} = 15,000-25,000 \, \mathrm{K}$ and electron densities $n_\mathrm{e} = 10^1-10^5 \, \mathrm{cm^{-3}}$. GS-NDG-9422 presents $\hdhb$ and $\hghb$ consistent with the Case B values, while the $\hahb$ is tentatively lower than the Case B value at the $>1\sigma$ level. RXCJ2248-ID shows $\hdhb$ and $\hghb$ larger than those of Case B at the $\sim 1.5\sigma$ and $\sim4\sigma$ levels, respectively, whereas the $\hahb$ is smaller than the Case B value at the $\sim8\sigma$ level. These deviations are opposite from the effect of the dust extinction. We therefore cannot explain these Balmer line ratios by the combination of Case B and dust extinction. 

\begin{figure*}
    \centering
    \includegraphics[width=0.9\linewidth]{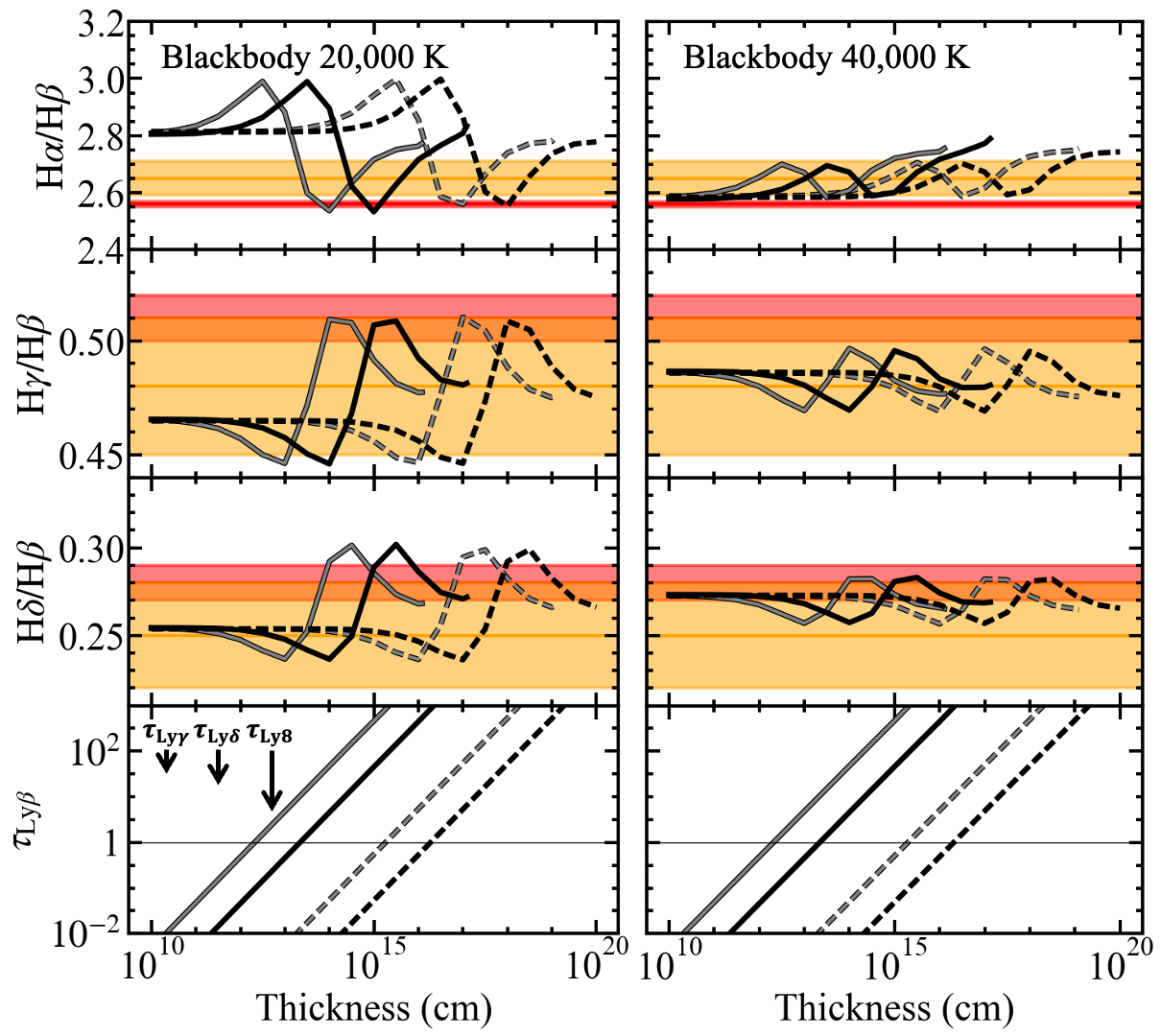}
    \caption{From top to bottom, $\hahb$, $\hghb$, $\hdhb$, and $\tau_{\mathrm{Ly}\beta}$ as a function of thickness of the nebula. The left and right panels correspond to the blackbody temperatures of $20,000 \, \mathrm{K}$ and $40,000 \, \mathrm{K}$, respectively. The dased and solid lines represent the values calculated at $n_\mathrm{e} = 10^2$ and $10^5 \, \mathrm{cm^{-3}}$, respectively. The gray and black lines indicate the values calculated at $\log U = -2$ and $-1$, respectively. $\tau_{\mathrm{Ly}\gamma}$, $\tau_{\mathrm{Ly}\delta}$, and $\tau_{\mathrm{Ly}8}$ are lower than $\tau_{\mathrm{Ly}\beta}$ by amounts corresponding to the length of the arrows in the bottom-left panel. The orange (red) line and shaded region denote the observed Balmer line ratios and their uncertainties of GS-NDG-9422 (RXCJ2248-ID), respectively. The horizontal lines in the bottom panels denote $\tau_\lambda = 1$, above which the line becomes optically thick. 
    }
    \label{fig:nebula_structure}
\end{figure*}

\subsection{
Optical Depth Smaller than the Case B Recombination
} \label{sec:hii}
Here we investigate the scenario with optically-thin Lyman series lines, which have smaller optical depths than those assumed in the Case B recombination.
This physical condition is realized in the density-bounded nebulae.
We simulate the density-bounded nebulae using version 22.02 of \textsc{Cloudy} \citep{Chatzikos+2023}. We vary the stopping thickness of the calculation from $10^{10}$ to $10^{20} \, \mathrm{cm}$. We change hydrogen densities from $10^2$ to $10^5 \, \mathrm{cm^{-3}}$, which cover the electron densities of $\sim10^3$ and $\sim10^5 \, \mathrm{cm^{-3}}$ for GS-NDG-9422 and RXCJ2248-ID reported by \cite{Cameron+2023} and \cite{Topping+2024}, respectively. We also vary ionization parameter $U$ from $\log U = -2$ to $-1$, motivated by the high ionization parameters in GS-NDG-9422 and RXCJ2248-ID implied by \cite{Cameron+2023} and \cite{Topping+2024}, respectively. For a shape of incident radiation, we choose blackbody with temperatures of 20,000 and 40,000 K, which are the typical temperatures for OB stars producing ionizing photons. Since Cloudy cannot assume a complex geometry, we assume a simple plane-parallel geometry. We fix chemical abundances to 10\% $Z_\odot$, which is comparable to the observed metallicities for GS-NDG-9422 and RXCJ2248-ID. The other parameters are summarized in Table
\ref{tab:cloudy_params}. 
The calculated Balmer line ratios and optical depths of Lyman series as a function of thickness of the nebula are shown in Figure \ref{fig:nebula_structure}.  
Figure \ref{fig:nebula_structure} indicates that the $\hahb$ ratios follow N-shaped changes as optical depths of Lyman series increase. This may be explained as follows (For detail, see Figure 4.3 of \citealt{Osterbrock+2006}). First, as the thickness increases, the optical depth of Ly$\beta$ ($\tau_{\mathrm{Ly}\beta}$; hereafter $\tau_\lambda$ refers to an optical depth of a line $\lambda$) approaches to $1$. This leads $\mathrm{Ly}\beta$ to be optically thick and a conversion of $\mathrm{Ly}\beta$ to $\ha$ occurs. When $\tau_{\mathrm{Ly}\gamma}$ approaches to 1,  
a conversion of $\mathrm{Ly}\gamma$ to $\hb$ occurs, which lowers the $\hahb$ ratio.
Conversions of higher order Lyman lines take place similarly for large thickness. When the optical depth around the one of Ly8 (transition from a principal quantum number of 8 to 1) is larger than 1, $\hahb$ gradually approaches to the Case B value. For $\hghb$, the first effect of increasing the thickness is the conversion of Ly$\gamma$ to $\hb$ due to $\tau_{\mathrm{Ly}\gamma}$ approaching to 1, which lowers the $\hghb$ ratio. As $\tau_{\mathrm{Ly}\delta}$ approaches to 1, a conversion of Ly$\delta$ to $\hg$ occurs, which leads to the increase of $\hghb$. Further increase of the thickness makes the higher order Lyman lines optically thick. When $\tau_{\mathrm{Ly}8}$ is above 1, $\hghb$ approaches to the Case B value. The $\hdhb$ ratios can also be explained similarly. 
For example, in the case of the blackbody temperature of $20,000 \, \mathrm{K}$, $\log n = 2$, and $\log U = -1$ (the gray solid lines in the left panels), the Balmer line ratios of GS-NDG-9422 and RXCJ2248-ID are reproduced when the thickness is $\sim10^{13.4}$ and $\sim10^{13.8} \, \mathrm{cm}$, respectively, where $\tau_{\mathrm{Ly}8}$ is $\sim1$. Fine-tuning of the thickness is thus required to reproduce the Balmer line ratios of these galaxies. For the other $U$ and $n$, the observed Balmer line ratios are reproduced at different thicknesses, while the necessity of the fine-tuning does not change.
In the case of the blackbody temperature of $40,000 \, \mathrm{K}$, the values of Balmer line ratios are different from those of $20,000 \, \mathrm{K}$ because of the temperature dependence of line emissivities. The Balmer line ratios of GS-NDG-9422 are reproduced within the wide ranges of thickness. However, for RXCJ2248-ID, $\hahb$ and $\hghb$ are not reproduced within the 1$\sigma$ level at any of the thickness, ionization parameter and density, while $\hdhb$ are reproduced for $\tau_{\mathrm{Ly}\beta} < -1$ or $\tau_{\mathrm{Ly}8} \sim 0.5$. We further investigate Balmer line ratios for other ranges of parameters (blackbody temperatures from $20,000$ to $50,000 \, \mathrm{K}$, ionization parameters from $\log U = -2$ to $0$, and densities from $\log n = 2$ to $5$). The calculated Balmer line ratios show the similar trends to those of the parameter sets discussed above.

Since the Balmer line ratios observed in GS-NDG-9422 and RXCJ2248-ID are determined with the small errors, only very narrow ranges of thickness are allowed to explain the Balmer line ratios. Fine-tuning of the thickness is thus necessary at given $U$, $n$, and blackbody temperature, and such situations are realized with a very low probability.

Recently \cite{McClymont+2024} report the Balmer decrement anomalies found in JADES galaxies. \cite{McClymont+2024} compare the observed Balmer line ratios with the Cloudy modeling of optically-thin nebulae, and find that the wide range of Balmer line ratios can be reproduced by varying metallicities from $10^{-3} Z_\odot$ to $10^{0} Z_\odot$. Although it is unclear whether metallicities of these galaxies and their modeling are consistent, \cite{McClymont+2024} claim that the metallicity is an important parameter to reproduce the Balmer decrement anomalies.

\subsection{
Optical Depth Larger than the Case B Recombination
} \label{sec:optical_depth} 
We investigate the scenario with optically-thick Balmer series lines, which have larger optical depths than those assumed in the Case B recombination.
We first investigate a simple optically-thick ionized nebula similar to those of \citet{Yano+2021}, who worked on nearby ultraluminous infrared galaxies for the Brackett lines, by \textsc{Cloudy} modeling similar to those in Section \ref{sec:hii}. However, we find no solutions explaining the Balmer decrement anomalies of GS-NDG-9422 and RXCJ2248-ID with any physical parameters in optically-thick ionized nebula. 

Instead, we explore the possibility that the ionization-bounded nebulae are observed through the thick excited neutral hydrogen gas. In other words, ionized nebulae are partly (or fully) surrounded by thick excited H \textsc{i} gas. 

\subsubsection{Analytic Calculations of Optical Depths} \label{sec:analytic_calculations}
The optical depth of a transition $\mathcal{N}' \rightarrow \mathcal{N}$, where $\mathcal{N}'$ and $\mathcal{N}$ are principal quantum numbers of a hydrogen atom that satisfy $\mathcal{N}' > \mathcal{N}$, is given by

\begin{equation} \label{eq:tau}
    \tau_{\mathcal{N}', \mathcal{N}} = \alpha_{\mathcal{N}, \mathcal{N}'} N_\mathcal{N},
\end{equation}
where $\alpha_{\mathcal{N}, \mathcal{N}'}$ is an absorption cross section of a transition $\mathcal{N} \rightarrow \mathcal{N}'$ and $N_\mathcal{N}$ is a column density of hydrogen atoms at the principal quantum number $\mathcal{N}$
\footnote{In this section, we follow the mathematical notations of \cite{Yano+2021}}.
Assuming that the line velocity profile is a Gaussian function, $\alpha_{\mathcal{N}, \mathcal{N}'}$ is given by

\begin{equation} \label{eq:alpha}
    \alpha_{\mathcal{N}, \mathcal{N}'} = \frac{hc}{4\pi^{3/2}} \frac{B_{\mathcal{N}, \mathcal{N}'}}{v_\mathrm{Dop}},
\end{equation}
where $h$ is the Planck constant, $c$ is the speed of light, $B_{\mathcal{N}, \mathcal{N}'}$ is a Einstein B coefficient of the transition $\mathcal{N} \rightarrow \mathcal{N}'$, and $v_\mathrm{Dop}$ is a Doppler velocity half width. If the velocity of the gas is determined by thermal motion, the Doppler velocity half width is expressed as $v_\mathrm{Dop} = \sqrt{2kT/m_\mathrm{H}}$, where $k$ is the Boltzmann constant, $T$ is the temperature of the gas, and $m_\mathrm{H}$ is the mass of the hydrogen atom. For $T \sim 20,000 \, \mathrm{K}$, which is approxmately the electron temperatures of GS-NDG-9422 and RXCJ2248-ID, we obtain $v_\mathrm{Dop} \sim 18 \, \mathrm{km \, s^{-1}}$. We assume the same $v_\mathrm{Dop}$ also for the neutral gas (although the absolute values of the optical depths depend on the assumption of $v_\mathrm{Dop}$, this ambiguity does not qualitatively change the following discussion). Substituting Equation (\ref{eq:alpha}) into Equation (\ref{eq:tau}) and using the Einstein coefficients taken from \cite{Wiese+2009}, we obtain

\begin{equation}\label{eq:taus}
\begin{split}
    \tau_{\mathrm{H}\alpha} &\sim 1.0 \left( \frac{N_2}{2.9\times10^{13} \, \mathrm{cm^{-3}}}\right) \left( \frac{v_\mathrm{Dop}}{18 \, \mathrm{km \, s^{-1}}}\right)^{-1} \\
    \tau_{\mathrm{H}\beta} &\sim 0.14 \left(\frac{N_2}{2.9\times10^{13} \, \mathrm{cm^{-3}}}\right) \left(\frac{v_\mathrm{Dop}}{18 \, \mathrm{km \, s^{-1}}}\right)^{-1} \\
    \tau_{\mathrm{H}\gamma} &\sim 0.05 \left(\frac{N_2}{2.9\times10^{13} \, \mathrm{cm^{-3}}}\right) \left(\frac{v_\mathrm{Dop}}{18 \, \mathrm{km \, s^{-1}}}\right)^{-1} \\
    \tau_{\mathrm{H}\delta} &\sim 0.02 \left(\frac{N_2}{2.9\times10^{13} \, \mathrm{cm^{-3}}}\right) \left(\frac{v_\mathrm{Dop}}{18 \, \mathrm{km \, s^{-1}}}\right)^{-1},
\end{split}
\end{equation}
We find that if $v_\mathrm{Dop} = 18 \, \mathrm{km \, s^{-1}}$ is assumed, $N_2$ as large as $2.9 \times 10^{13} \, \mathrm{cm^{-2}}$ makes $\ha$ optically thick, while the other Balmer lines remain optically thin. In this case, observed Balmer line ratios are calculated as

\begin{equation}
    \frac{\mathrm{H}n}{\mathrm{H}\beta}_\mathrm{obs} = \frac{\mathrm{H}n}{\mathrm{H}\beta}_\mathrm{int} \times \exp\left({-\tau_{\mathrm{H}n}}+\tau_{\mathrm{H}\beta}\right),
\end{equation}
where $n$ represents $\alpha$, $\gamma$, or $\delta$. We calculate $(\mathrm{H}n/\mathrm{H}\beta)_\mathrm{int}$ using PyNeb and optical depths are taken from Equation \eqref{eq:taus}, where we assume $v_\mathrm{Dop} = 18 \, \mathrm{km\,s^{-1}}$. 

Although the absorption is generally followed by re-emission, we do not consider the effect of re-emission. Such a situation is achieved by an asymmetric geometry, as suggested by \cite{Scarlata+2024}. \cite{Scarlata+2024} have claimed that, if the gas has a flattened geometry and an edge-on orientation, the re-emitted photon is easily scattered out of the line of sight, which makes the re-emission of the Balmer line photon negligible (see Figure 5 of \citealt{Scarlata+2024}).

\begin{figure*}
    \centering
    \includegraphics[width=0.8\linewidth]{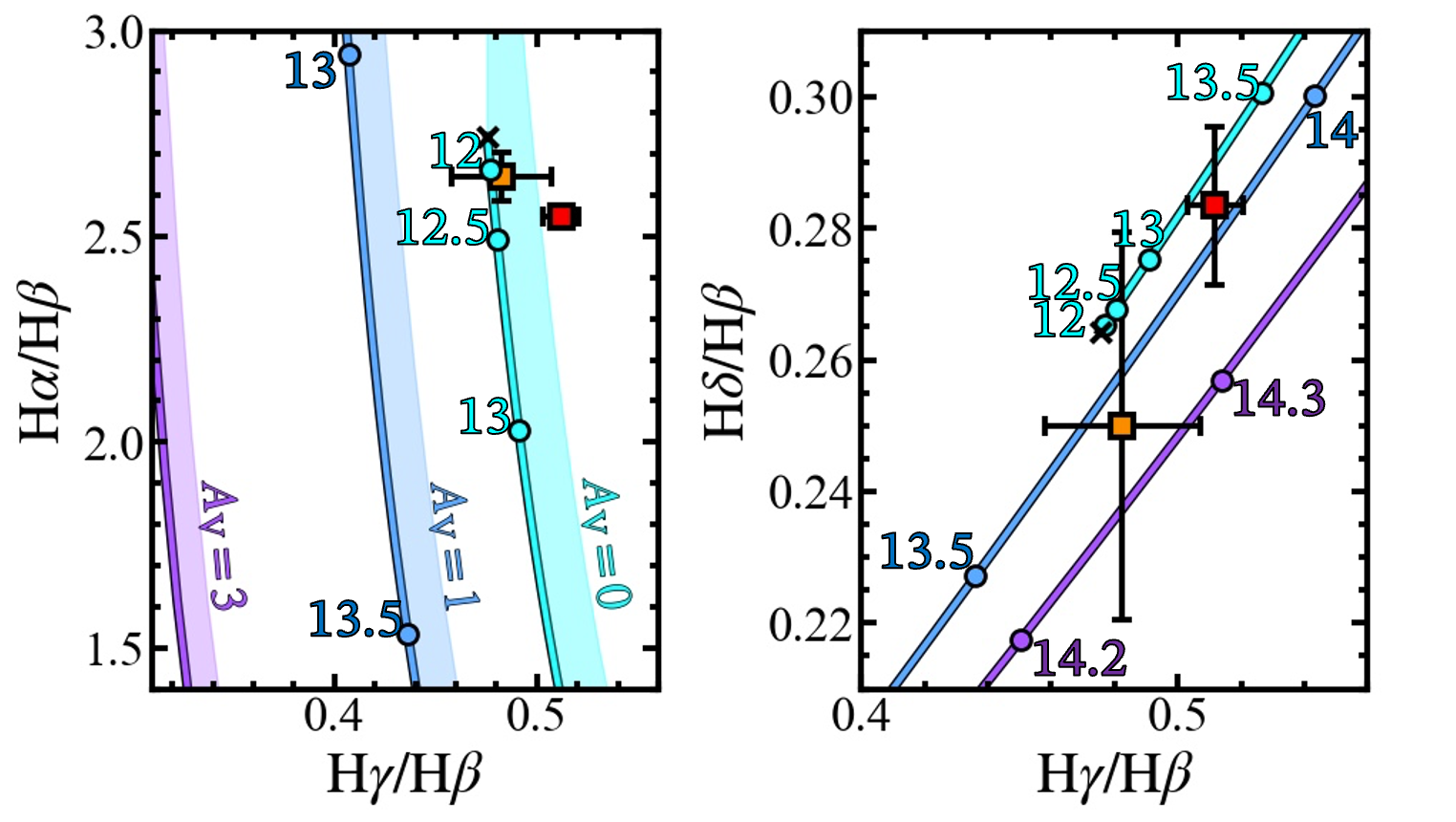}
    \caption{Comparisons between the observed Balmer line ratios and theoretical Balmer line ratios including the effects of the optical depths of the Balmer lines. The squares are the same as Figure \ref{fig:Balmer_decrements}. The black crosses indicate the Case B values for $T_\mathrm{e} = 20,000 \, \mathrm{K}$ and $n_\mathrm{e} = 10^3 \, \mathrm{cm^{-3}}$. The cyan, blue, and purple lines and circles denote the Balmer line ratios including the effects of the optical depths with $A_\mathrm{V} = 0$, $1$, and $3$, respectively. The numbers shown beside the circles represent the values of $\log N_2$. The cyan, blue, and purple shaded region in the left panel indicate the regions that can be reproduced by the interstellar shock \citep{Shull+1979} in the case of $A_\mathrm{V} = 0, 1$, and $3$, respectively.}
    \label{fig:optical_depth}
\end{figure*}

In Figure \ref{fig:optical_depth}, the Balmer line ratios attenuated by neutral hydrogen are plotted for several $N_2$ and dust extinction $A_\mathrm{V}$ values. 
Note that we treat $N_2$ and $A_\mathrm{V}$ as the independent free parameters, because the gas-to-dust ratio is unknown for the high-$z$ metal-poor galaxies. 
Here an SMC extinction law is assumed \citep{Gordon+2003}. RXCJ2248-ID shows broad components in $\ha$ and [O \textsc{iii}] $\lambda$5007 with FWHMs of 607 and 1258 $\mathrm{km \, s^{-1}}$, respectively, which implies existence of interstellar shock. The effect of interstellar shock thus needs to be incorporated in the model. In Figure \ref{fig:optical_depth}, the cyan, blue, and purple shaded regions present the effect of interstellar shock on $\hahb$ suggested by \cite{Shull+1979}, who present that interstellar shock increase $\hahb$ by up to $\sim1.6$ times. 
Figure \ref{fig:optical_depth} presents that
larger $N_2$ leads to smaller $\hahb$ and larger $\hghb$ and $\hdhb$. This is because the line with the longer wavelength has larger probability of absorption (Equation \ref{eq:taus}). Comparing the models with the observational results, we find that Balmer line ratios of GS-NDG-9422 and RXCJ2248-ID are reproduced for $\log N_2 \sim 12 - 13$ and $A_\mathrm{V} \sim 0$ within the $\sim1\sigma$ levels. 
%

\subsubsection{MCMC Analyses} \label{sec:mcmc}
\begin{figure*}
    \centering
    \includegraphics[width=1\linewidth]{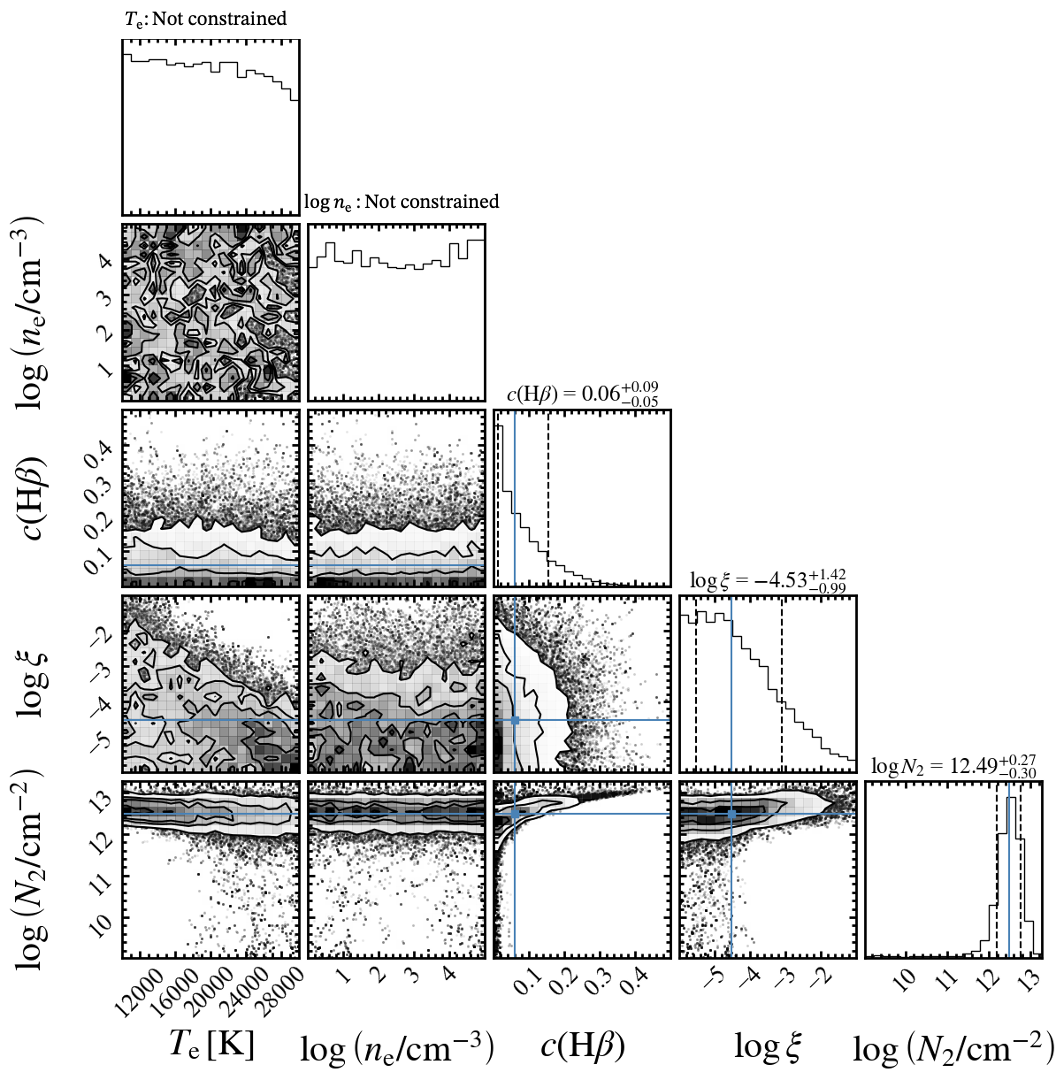}
    \caption{Probability distributions for GS-NDG-9422 recovered with modified \textsc{ymcmc}. The diagonal panels show the 1D probability distribution for each parameter. The off-diagonal panels present the 2D contours. The blue solid and black dashed lines indicate the median and the 16th and 84th percentiles of the probability distributions, respectively.}
    \label{fig:mcmc_cameron}
\end{figure*}

\begin{figure*}
    \centering
    \includegraphics[width=1\linewidth]{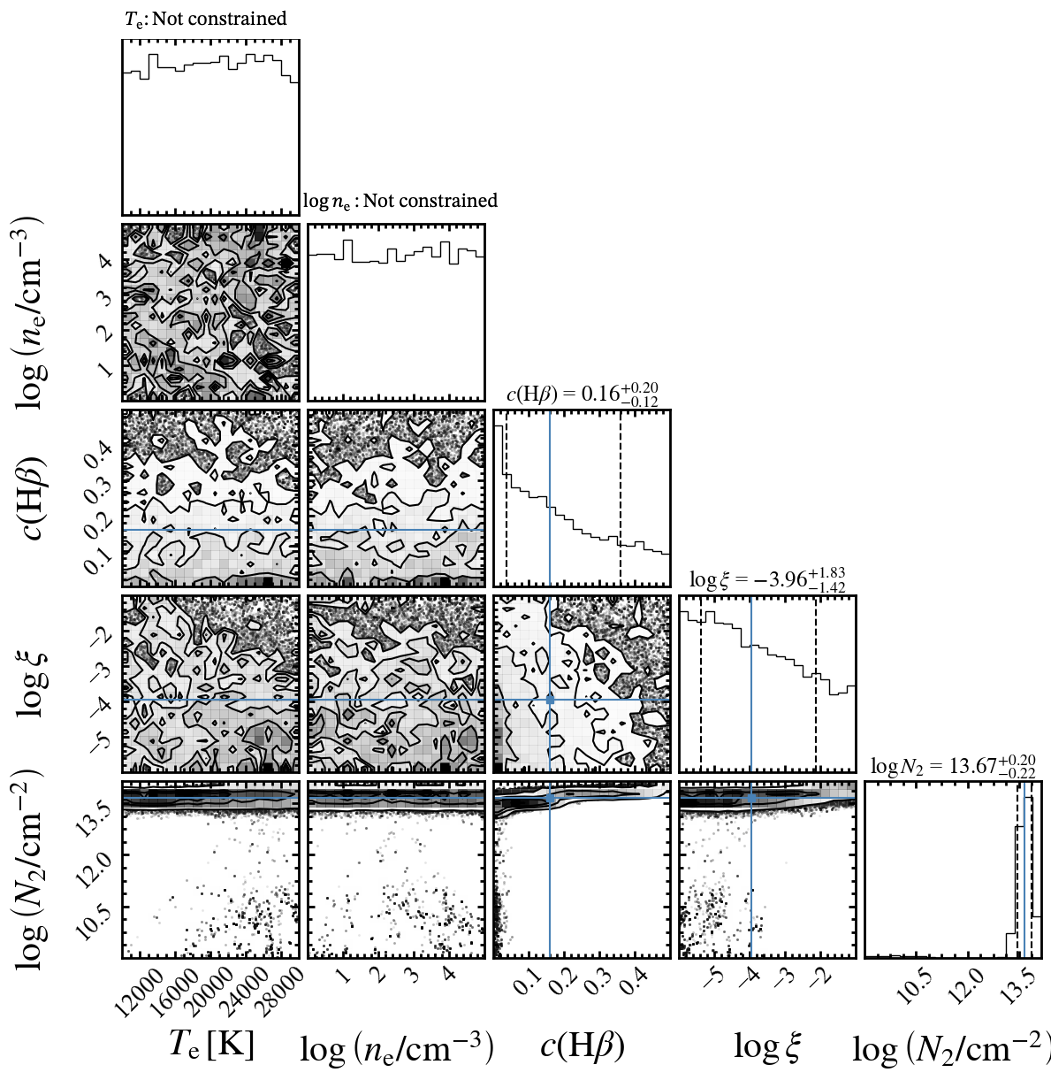}
    \caption{Same as Figure \ref{fig:mcmc_cameron}, but for RXCJ2248-ID.}
    \label{fig:mcmc_topping}
\end{figure*}

To estimate $N_2$ self-consistently with other physical parameters, we use Markov Chain Monte Carlo (MCMC) algorithm. We utilized \textsc{ymcmc} code developed by \cite{Hsyu+2020}. Although \textsc{ymcmc} have originally been developed to derive helium abundances of galaxies, we rewrite the code to model hydrogen line fluxes with the optical depth effects. Hereafter this code is referred to as modified \textsc{ymcmc}. We set 5 free parameters of $T_\mathrm{e}$, $n_\mathrm{e}$, a reddening parameter $c(\mathrm{H}\beta)$, a ratio of neutral to ionized hydrogen density $\xi$, and $N_2$. At each step of the MCMC, model flux ratios are calculated by

\begin{multline}
    \frac{F(\lambda)}{F(\mathrm{H}\beta)}_\mathrm{mod} = \frac{E(\lambda)}{E(\mathrm{H}\beta)} \frac{1+\frac{C}{R}(\lambda)}{1+\frac{C}{R}(\mathrm{H}\beta)} 10^{-f(\lambda)\,c(\mathrm{H}\beta)} \\
    \times \exp{\left(-\tau_\lambda+\tau_{\mathrm{H}\beta}\right)},
\end{multline}
where $F(\lambda)$, $E(\lambda)$, $C/R(\lambda)$, and $f(\lambda)$ represent the flux, emissivity, collisional-to-recombination coefficient, and reddening law, respectively. See \cite{Hsyu+2020} for the details of these parameters. We minimize a log-likelihood function given by

\begin{equation}
    \log\left(\mathcal{L}\right) = \sum_{\lambda}\frac{\left(\frac{F(\lambda)}{F\left(\mathrm{H}\beta\right)}_\mathrm{obs}-\frac{F(\lambda)}{F\left(\mathrm{H}\beta\right)}_\mathrm{mod}\right)^2}{\sigma(\lambda)^2},
\end{equation}
where the subscript `obs' denotes the observed value and $\sigma(\lambda)$ represents the $1\sigma$ uncertainty of the observed flux of the line $\lambda$. We use flat priors of

\begin{equation}\label{eq:prior}
\begin{gathered}
    10,000 \leq T_\mathrm{e} \leq 30,000\\
    1 \leq \log_{10}n_\mathrm{e} \leq 5\\
    0 \leq c(\mathrm{H}\beta) \leq 0.5\\
    -6 \leq \log_{10} \xi \leq -1 \\
    9 \leq \log_{10} N_2 \leq 15.
\end{gathered}
\end{equation}
We do not use $\ha$ for RXCJ2248-ID to avoid contamination by the interstellar shock. 
The probability distributions for GS-NDG-9422 and RXCJ2248-ID are presented in Figure \ref{fig:mcmc_cameron} and \ref{fig:mcmc_topping}, respectively. The best-fit parameters are shown in Table \ref{tab:bestfit}. We obtain $\log \left(N_2 / \mathrm{cm^{-2}} \right) = 12.5^{+0.3}_{-0.3}$ and $13.7^{+0.2}_{-0.2}$ for GS-NDG-9422 and RXCJ2248-ID, respectively. The probability distributions for $c(\mathrm{H}\beta)$ prefer values close to 0, implying dust-poor conditions. These results are consistent with those seen in Figure \ref{fig:optical_depth}. The electron temperatures and densities are not constrained because the Balmer line ratios are only weakly dependent on these parameters. In Appendix \ref{sec:appendix}, we apply Gaussian priors for $T_\mathrm{e}$ and $n_\mathrm{e}$ that are determined by the line diagnostics of the previous studies, and obtain constraints on these parameters. Whether the Gaussian priors are applied or not, our main results do not change.

\begin{deluxetable*}{cccccc}
    \centering
    \tablecaption{Best-fit parameters recovered with modified \textsc{ymcmc}}
    \tablehead{
    \colhead{ID} & \colhead{$T_\mathrm{e}$} & \colhead{$\log \left(n_\mathrm{e}/\mathrm{cm^{-3}}\right)$} & \colhead{$c\left(\mathrm{H}\beta\right)$} & \colhead{$\log \xi$} & \colhead{$\log \left(N_2/\mathrm{cm^{-2}}\right)$} \\
    \colhead{} & \colhead{[K]} & \colhead{} & \colhead{} & \colhead{} & \colhead{}
    }
    \startdata
    GS-NDG-9422 & \dots & \dots & $0.06^{+0.09}_{-0.05}$ & $-4.53^{+1.42}_{-0.99}$ & $12.49^{+0.27}_{-0.30}$ \\
    RXCJ2248-ID & \dots & \dots & $0.18^{+0.20}_{-0.12}$ & $-3.96^{+1.83}_{-1.42}$ & $13.67^{+0.20}_{-0.22}$ \\
    \enddata
    \tablecomments{Median and the 16th and 84th percentiles of the probability distributions are shown. Values for $T_\mathrm{e}$ and $n_\mathrm{e}$ are not presented because the obtained distributions are almost constant.}
    \label{tab:bestfit}
\end{deluxetable*}

\section{Discussion} \label{sec:discussion}
In the case where the Lyman lines are optically thin discussed in Section \ref{sec:hii}, the nebula is density-bounded. This indicates the strong ionizing radiation in the high-$z$ galaxies, which is in line with the large O32 value reported in RXCJ2248-ID \citep{Topping+2024}. Also in this case, the nebula can provide a large number of $\mathrm{Ly}\alpha$ photons, which is compatible with the strong $\mathrm{Ly}\alpha$ emission observed in GS-NDG-9422 \citep{Cameron+2023} and RXCJ2248-ID \citep{Mainali+2017}. The density-bounded nebulae can provide ionizing photon with high escape fraction, which may play a key role for cosmic reionization \citep{Nakajima+2014, Naidu+2022}. In the density-bounded situation, however, it is difficult to explain the strong two-photon emission observed in GS-NDG-9422 \citep{Cameron+2023}. The two-photon emission occurs in the forbidden $2s\rightarrow 1s$ transition. The large $N_2$ value is needed to produce the strong two-photon emission, which is difficult to achieve in the density-bounded nebula due to the low column density. An existence of a damped Ly$\alpha$ (DLA) system with 30\% leakage is also suggested as an alternative explanation for the spectral feature of GS-NDG-9422, while \cite{Cameron+2023} regard it as unlikely. Although the density-bounded nebulae can coexist with the DLA, scattering or absorption of the Balmer line photons by DLA are likely to occur, which change the observed Balmer line ratios. The condition with optically-thin Lyman lines does not contradict the dense situation observed in RXCJ2248-ID \citep{Topping+2024}.

In the case where the Balmer lines are optically-thick discussed in Section \ref{sec:optical_depth}, $N_2$ needs to be large to make Balmer lines optically thick. To increase $N_2$ in the neutral hydrogen gas, a large number of hydrogen atoms at the ground state need to be excited to the second energy level. Although a physical origin of this excessive excitation is not constrained, our result may suggest that supernovae or accretion from the large scale structure to a halo kinematically cause collisional excitation of the hydrogen gas in the high-$z$ galaxies \citep{Dekel+2009, Yajima+2015, Shimoda+2019}. Our result of large $N_2$ qualitatively compatible with the strong two-photon emission observed in GS-NDG-9422 \citep{Cameron+2023}, because a large number of hydrogen atoms at the $2s$ state produce strong two-photon emission via the $2s \rightarrow 1s$ transition. The neutral gas with optically-thick Balmer lines can coexist with the dense ionized regions that are reported in RXCJ2248-ID \citep{Topping+2024}.

The Balmer line ratios play a key role to determine the amount of dust extinction. If the estimate of dust extinction is not correct, other nebular physical quantities such as the electron temperature and chemical abundances are also wrong. It is thus important to understand the origin of the Balmer decrement anomalies. 

\section{Summary} \label{sec:summary}
In this paper, the physical mechanisms that cause the observed Balmer decrement anomalies are investigated. Our main findings are summarized below:

\begin{enumerate}
    \item GS-NDG-9422 and RXCJ2248-ID found with JWST at $z\sim6$ show $\hahb$ significantly smaller than the Case B values as reported by \citet{Cameron+2023} and \citet{Topping+2024}, respectively. RXCJ2248-ID also presents $\hghb$ and $\hdhb$ significantly larger than the Case B values. The observed Balmer line ratios are not explained by the combination of the dust extinction and Case B, since the observed deviation from the Case B values are opposite to the effect of the dust extinction. Changing the electron temperature and density also does not reproduce the observed Balmer line ratios.

    \item Using \textsc{Cloudy}, we model the density-bounded nebula where the Lyman lines are optically thin. We calculate the Balmer line ratios and the optical depth of Ly$\alpha$ with several stopping thicknesses. We find that the optical depths of Lyman series changes the Balmer line ratios due to the conversions of the Lyman lines to Balmer lines. We find that when the nebula is optically thick only up to Ly$\gamma-$Ly8, the observed Balmer line ratios are reproduced. However, this situation requires fine-tuning of the thickness of the nebula. The density-bounded situation suggests the strong ionizing radiation, which qualitatively agree with the large O32 observed in RXCJ2248-ID. The nebula that is optically-thin to Lyman lines is also compatible with the strong Ly$\alpha$ emission observed in GS-NDG-9422. The density-bounded nebulae supply ionizing photon with high escape fraction, which may contribute to the cosmic reionization. However, the density-bounded situation is difficult to explain the strong two-photon continuum observed in GS-NDG-9422. 
    
    \item We also model the neutral hydrogen gas that surround the ionized region and is optically thick to the Balmer lines. We find that, if $N_2$ is larger than $\sim 10^{13} \, \mathrm{cm^{-2}}$, the Balmer lines become optically thick, and that the observed Balmer line ratios are reproduced. We develop nebular models with neutral gas that surround the ionized gas and have large $N_2$. Our nebular models extending the \textsc{ymcmc} models \citep{Hsyu+2020} suggest $\log \left(N_2 / \mathrm{cm^{-2}} \right) = 12.5^{+0.3}_{-0.3}$ and $13.7^{+0.2}_{-0.2}$ for GS-NDG-9422 and RXCJ2248-ID, respectively, as well as the dust-poor conditions. This results suggest that the excited H \textsc{i} clouds that partly (or fully) surround the ionized region may be the cause of the Balmer decrement anomalies. Although a physical process that enhances $N_2$ in the neutral gas around the ionized gas is unclear, there may be accretion or supernovae that cause excessive collisional excitation of the second energy level of hydrogen atoms. Our result of large $N_2$ for GS-NDG-9422 in line with the strong two-photon continuum suggested by \cite{Cameron+2023}. Because a large number of hydrogen atoms at the $2s$ state produce the two-photon emission via the $2s\rightarrow1s$ transition. The neutral hydrogen nebula with high $N_2$ can coexist the dense ionized nebula reported in RXCJ2248-ID \citep{Topping+2024}.
\end{enumerate}

\section*{acknowledgments}
We thank Michael W. Topping for providing the information about the $\ha$ flux reported by \cite{Topping+2024}. We are thankful to Alex J. Cameron and Kentaro Nagamine for helpful comments. This work use the results reported by Alex J. Cameron and Michael W. Topping. We acknowledge Alex J. Cameron and Michael W. Topping for publicly releasing their analysed data. This work is based on the observations with the NASA/ESA/CSA James Webb Space Telescope associated with programs of GTO-1210 (JADES) and GO-2478. We are grateful to GTO-1210 and GO-2478 teams led by Nora Luetzgendorf and Daniel P. Stark, respectively. This publication is based upon work supported by the World Premier International Research Center Initiative (WPI Initiative), MEXT, Japan, KAKENHI (20H00180, 21H04467, 21H04489, 21H04496, and 23H05441) through Japan Society for the Promotion of Science, and JST FOREST Program (JP-MJFR202Z). This work was supported by the joint research program of the Institute for Cosmic Ray Research (ICRR), University of Tokyo.

\vspace{5mm}

\software{Astropy \citep{astropy_2013, astropy_2018, Astropy+2022}, Cloudy \citep{Chatzikos+2023}, matplotlib \citep{Hunter+2007}, NumPy \citep{Harris+2020}, PyNeb \citep{Luridiana+2015}, SciPy \citep{Virtanen+2020}, and YMCMC \citep{Hsyu+2020}.
          }

\appendix

\section{Gaussian Priors for $T_\mathrm{\lowercase{e}}$ and $\lowercase{n_\mathrm{e}}$ in MCMC Analyses} \label{sec:appendix}
\begin{figure*}
    \centering
    \includegraphics[width=1\linewidth]{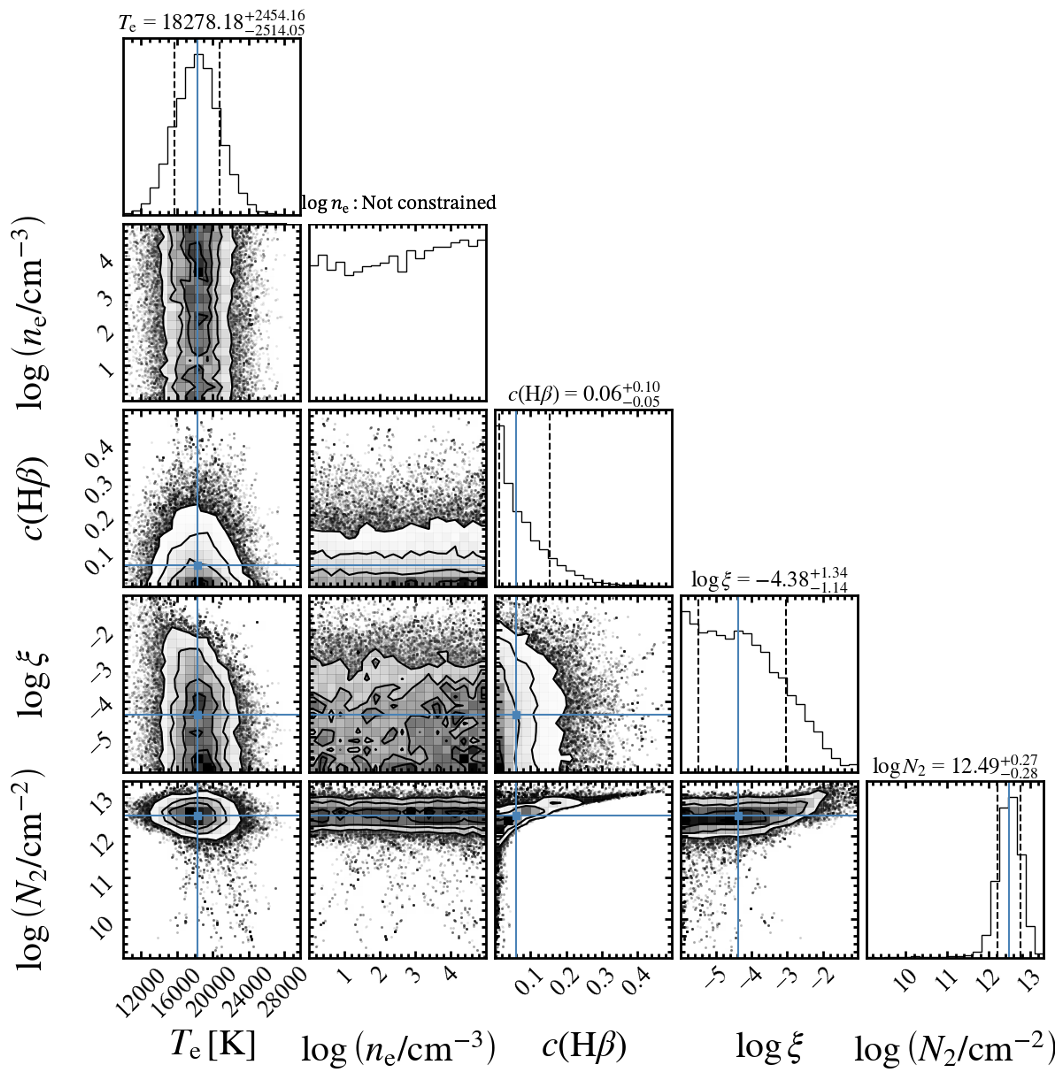}
    \caption{
    Same as Figure \ref{fig:mcmc_cameron}, but for GS-NDG-9422 with Gaussian priors for $T_\mathrm{e}$.
    }
    \label{fig:mcmc_cameron_2}
\end{figure*}

\begin{figure*}
    \centering
    \includegraphics[width=1\linewidth]{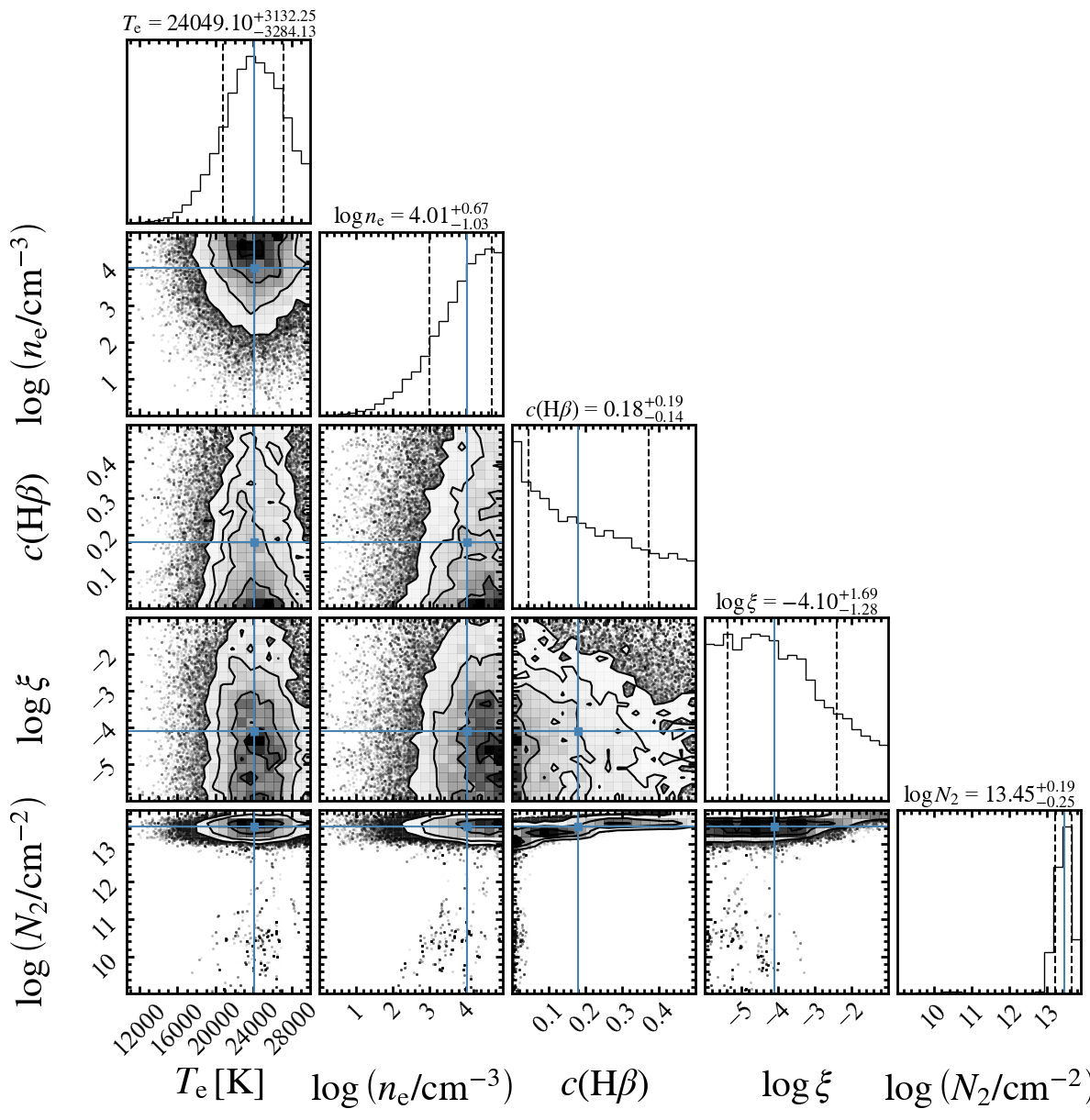}
    \caption{Same as Figure \ref{fig:mcmc_cameron}, but for RXCJ2248-ID with Gaussian priors for $T_\mathrm{e}$ and $n_\mathrm{e}$.}
    \label{fig:mcmc_topping_2}
\end{figure*}

The flat priors \eqref{eq:prior} give the almost constant probability distributions for $T_\mathrm{e}$ and $n_\mathrm{e}$, since the dependence of Balmer decrements on these parameters are weak. However, providing independent measurements of $T_\mathrm{e}$ and $n_\mathrm{e}$ may give physically more reasonable distributions. Here, using $T_\mathrm{e}$ and $n_\mathrm{e}$ derived from emission line diagnostics, we put weak Gaussian priors on these parameters. This procedure is proposed by \cite{Aver+2011}, who have used mock data and demonstrated that the input model parameters are recovered more accurately when the weak Gaussian prior is imposed.
Given the $T_\mathrm{e}$ and $n_\mathrm{e}$ values derived from the emission line diagnostics, log-likelihood is calculated as

\begin{equation}\label{eq:loglike_with_gaussian}
    \log\left(\mathcal{L}\right) = \sum_{\lambda}\frac{\left(\frac{F(\lambda)}{F\left(\mathrm{H}\beta\right)}_\mathrm{obs}-\frac{F(\lambda)}{F\left(\mathrm{H}\beta\right)}_\mathrm{mod}\right)^2}{\sigma(\lambda)^2} + \frac{(T_\mathrm{e, obs}-T_\mathrm{e, mod})^{2}}{2\sigma_{T_\mathrm{e}}^{2}} + \frac{(n_\mathrm{e, obs}-n_\mathrm{e, mod})^{2}}{2\sigma_{n_\mathrm{e}}^{2}},
\end{equation}
where $T_\mathrm{e, obs}$ and $n_\mathrm{e,obs}$ are the electron temperature and density derived from observation, respectively, and $T_\mathrm{e, mod}$ and $n_\mathrm{e, mod}$ are those of the model parameters. $\sigma_{T_\mathrm{e}}$ and $\sigma_{n_\mathrm{e}}$ are the standard deviations of the Gaussian distribution. Here we take $\sigma_{T_\mathrm{e}} $ and $\sigma_{n_\mathrm{e}}$ to be conservative values of $0.2 \, T_\mathrm{e, obs}$ and $0.4 \, n_\mathrm{e, obs}$, respectively, the former of which is proposed by \cite{Aver+2011}. The obtained probability distributions for GS-NDG-9422 and RXCJ2248-ID are presented in Figure \ref{fig:mcmc_cameron_2} and \ref{fig:mcmc_topping_2}, respectively. The $T_\mathrm{e, obs}$ and $n_\mathrm{e, obs}$ values and best-fit parameters are summarized in Table \ref{tab:bestfit_2}. Note that $n_\mathrm{e, obs}$ is not used for GS-NDG-9422 (the last term in right hand side of Equation \eqref{eq:loglike_with_gaussian} is omitted) because the only weak upper limit on $n_\mathrm{e}$ is provided \citep{Cameron+2023}. The best-fit $N_2$ values are $\sim 12$ and $\sim 13$ for GS-NDG-9422 and RXCJ2248-ID, respectively, indicating that imposing the weak Gaussian priors does not change our main result.

\begin{deluxetable*}{cccccc}
    \centering
    \tablecaption{Best-fit parameters recovered with modified \textsc{ymcmc}}
    \tablehead{
    \colhead{ID} & \colhead{$T_\mathrm{e}$} & \colhead{$\log \left(n_\mathrm{e}/\mathrm{cm^{-3}}\right)$} & \colhead{$c\left(\mathrm{H}\beta\right)$} & \colhead{$\log \xi$} & \colhead{$\log \left(N_2/\mathrm{cm^{-2}}\right)$} \\
    \colhead{} & \colhead{[K]} & \colhead{} & \colhead{} & \colhead{} & \colhead{}
    }
    \startdata
    GS-NDG-9422 & $18278^{+2454}_{-2514}$ & $\dots$ & $0.06^{+0.10}_{-0.05}$ & $-4.38^{+1.34}_{-1.14}$ & $12.49^{+0.27}_{-0.28}$ \\
    RXCJ2248-ID & $24049^{+3132}_{-3284}$ & $4.01^{+0.67}_{-1.03}$ & $0.18^{+0.19}_{-0.14}$ & $-4.10^{+1.69}_{-1.28}$ & $13.45^{+0.19}_{-0.25}$ \\
    \enddata
    \tablecomments{Same as Table \ref{tab:bestfit}, but with Gaussian priors for $T_\mathrm{e}$ and $n_\mathrm{e}$. For GS-NDG-9422, since $n_\mathrm{e}$ derived from line diagnostics does not exist, no constraint on $n_\mathrm{e}$ is obtained.}
    \label{tab:bestfit_2}
\end{deluxetable*}

\bibliography{reference}{}
\bibliographystyle{aasjournal}

\end{document}